\documentclass[prl,twocolumn,showpacs,superscriptaddress]{revtex4}

\usepackage{graphicx}
\usepackage{dcolumn}

\usepackage{eucal}
\usepackage[dvips]{epsfig}
\usepackage{amssymb}

\usepackage{amsmath,amsthm}

\newcommand{\be}{\begin{eqnarray}}
\newcommand{\ee}{\end{eqnarray}}
\newcommand{\bse}{\begin{subequations}}
\newcommand{\ese}{\end{subequations}}

\newcommand{\bnum}{\begin{enumerate}}
\newcommand{\enum}{\end{enumerate}}

\newcommand{\bit}{\begin{itemize}}
\newcommand{\eit}{\end{itemize}}

\newcommand{\bc}{\begin{cases}}
\newcommand{\ec}{\end{cases}}

\newcommand{\bpm}{\begin{pmatrix}}
\newcommand{\epm}{\end{pmatrix}}

\newcommand{\bvm}{\begin{vmatrix}}
\newcommand{\evm}{\end{vmatrix}}

\newcommand{\bs}{\boldsymbol}

\newcommand{\ovl}{\overline}

\newcommand{\gb}{\beta}

\newcommand{\gl}{\lambda}

\newcommand{\go}{\omega}

\newcommand{\Go}{\Omega}
\newcommand{\Gc}{\Gamma}

\newcommand{\Gd}{\Delta}

\newcommand{\Gl}{\Lambda}

\newcommand{\p}{\partial}

\newcommand{\lan}{\langle}
\newcommand{\ran}{\rangle}

\begin{document}
\title{Fluid dynamics of bacterial bulk turbulence}
\title{Fluid dynamics of bacterial turbulence}

%

\author{J\"orn Dunkel}
\affiliation{DAMTP, Centre for
Mathematical Sciences, University of Cambridge, Wilberforce Road, Cambridge CB3 0WA, UK}

\author{Sebastian Heidenreich}
\affiliation{Physikalisch-Technische Bundesanstalt, Abbestr. 2-12, 10587 Berlin, Germany}

\author{Knut Drescher}
\affiliation{Departments of Molecular Biology and Mechanical and Aerospace Engineering, Princeton University, Princeton, New Jersey 08544}

\author{Henricus H. Wensink}
\affiliation{Laboratoire de Physique des Solides, Universit\'{e} Paris-Sud 11 \& CNRS, B\^{a}timent 510, 91405 Orsay Cedex, France}

\author{Markus B\"ar}
\affiliation{Physikalisch-Technische Bundesanstalt, Abbestr. 2-12, 10587 Berlin, Germany}

\author{Raymond E. Goldstein}
\affiliation{DAMTP, Centre for
Mathematical Sciences, University of Cambridge, Wilberforce Road, Cambridge CB3 0WA, UK}

\date{\today}
\begin{abstract} 
Self-sustained turbulent structures have been observed in a wide range of living fluids, yet no quantitative theory exists
to explain their properties. We report experiments
on active turbulence in highly concentrated 3D suspensions of 
\textit{Bacillus subtilis} and compare them with a minimal fourth-order vector-field theory for incompressible 
bacterial dynamics.  Velocimetry of bacteria and surrounding fluid, determined by imaging cells and 
tracking colloidal tracers, yields consistent results for velocity statistics and correlations over two orders of magnitude in 
kinetic energy, revealing a decrease of fluid memory with increasing swimming activity and linear scaling between 
energy and enstrophy.  The best-fit model parameters allow for quantitative agreement with 
experimental data.
\end{abstract}

\pacs{87.10.-e,87.10.Ed,87.18.Hf}

 
\maketitle

A series of experiments over the last decade~\cite{2011Couzin,2009Parisi,1997Kessler,2004DoEtAl,2007SoEtAl,Swinney_bactclust,2012Wensink,2012Sokolov,2010Bausch,2012Sumino}
has shed light on generic ordering principles that appear to govern collective dynamics of living
matter~\cite{2009CoWe,2011KochSub,2012Vicsek,2010Ramaswamy,2012Marchetti_Review}, from  large-scale 
animal swarming~\cite{2011Couzin,2009Parisi} to meso-scale turbulence 
in microbial 
suspensions~\cite{1997Kessler,2004DoEtAl,2007SoEtAl,Swinney_bactclust,2012Wensink,2012Sokolov} 
and micro-scale 
self-organization in motility assays~\cite{2010Bausch,2012Sumino}. Although very different in 
size and composition, 
these systems are often jointly termed \lq active\rq\space fluids, for which there is now a range of 
continuum theories~\cite{2004Kruse_PRL,2010Ramaswamy,2012Marchetti_Review,2008Wolgemuth,2011KochSub,1998TonerTu_PRE,2009BaMa_PNAS,2008SaintillanShelley,2012Peshkov,2013Lauga}. 
From these have come important qualitative insights into instability
mechanisms~\cite{2004Kruse_PRL,2010Ramaswamy,2012Vicsek,2012Marchetti_Review,2008SaintillanShelley,2013Grossmann} 
driving dynamical pattern formation, but a quantitative picture remains inchoate; even for the 
simplest active (e.g., bacterial or algal) suspensions uncertainty remains about which hydrodynamic equations 
and transport coefficients~\cite{2009Rafai,2009SoAr} provide an adequate minimal description, due in large part to the 
inability of existing data to constrain the manifold parameters in these models.
One approach to remedy this problem is to characterize collective dynamics as in
high Reynolds number fluid turbulence,  in terms of kinetic energy, enstrophy and spatio-temporal correlation functions, and to compare with an appropriate 
long-wavelength theory (i.e. Navier-Stokes-type equations).  We present such an analysis here, 
measuring collective behavior in dense suspensions of the bacterium {\it Bacillus subtilis} in comparison to
predictions of a (fourth-order) continuum model for bacterial flow~\cite{2012DuEtAl,2012Wensink}.

Previous experimental studies of bacterial suspensions in open
droplets~\cite{1997Kessler,2004DoEtAl,2005Tuval_PNAS,2011Cisneros_PRE},
freestanding films~\cite{2007SoEtAl,2009SoAr,2012Lin,2012Sokolov},
on surfaces~\cite{Swinney_bactclust,2009Swinney_EPL,2012Peruani}, or quasi-2D microfluidic chambers~\cite{2012Wensink}
focused separately on the bacterial and fluid components, leaving uncertain how accurately passive tracers~\cite{2000WuLi,2011Japan} reflect collective bacterial dynamics.  The experiments reported here, 
performed in closed 3D microfluidic chambers, allowed near-simultaneous measurements of cell and tracer motion,
and exploit a natural reduction in bacterial swimming activity due 
to oxygen depletion~\cite{2005Tuval_PNAS,2009Douarche,2012Sokolov} to obtain data spanning two 
orders of magnitude in fluid kinetic energy.  Combined with extensive 3D numerical simulations of the model, this data 
allows robust parameter estimates.  Quantitative agreement between experiment and 
theory suggests that this model presents a viable generalization of the Navier-Stokes equations to incompressible active fluids. 

\begin{figure}[t]
\includegraphics[width=1.0\columnwidth]{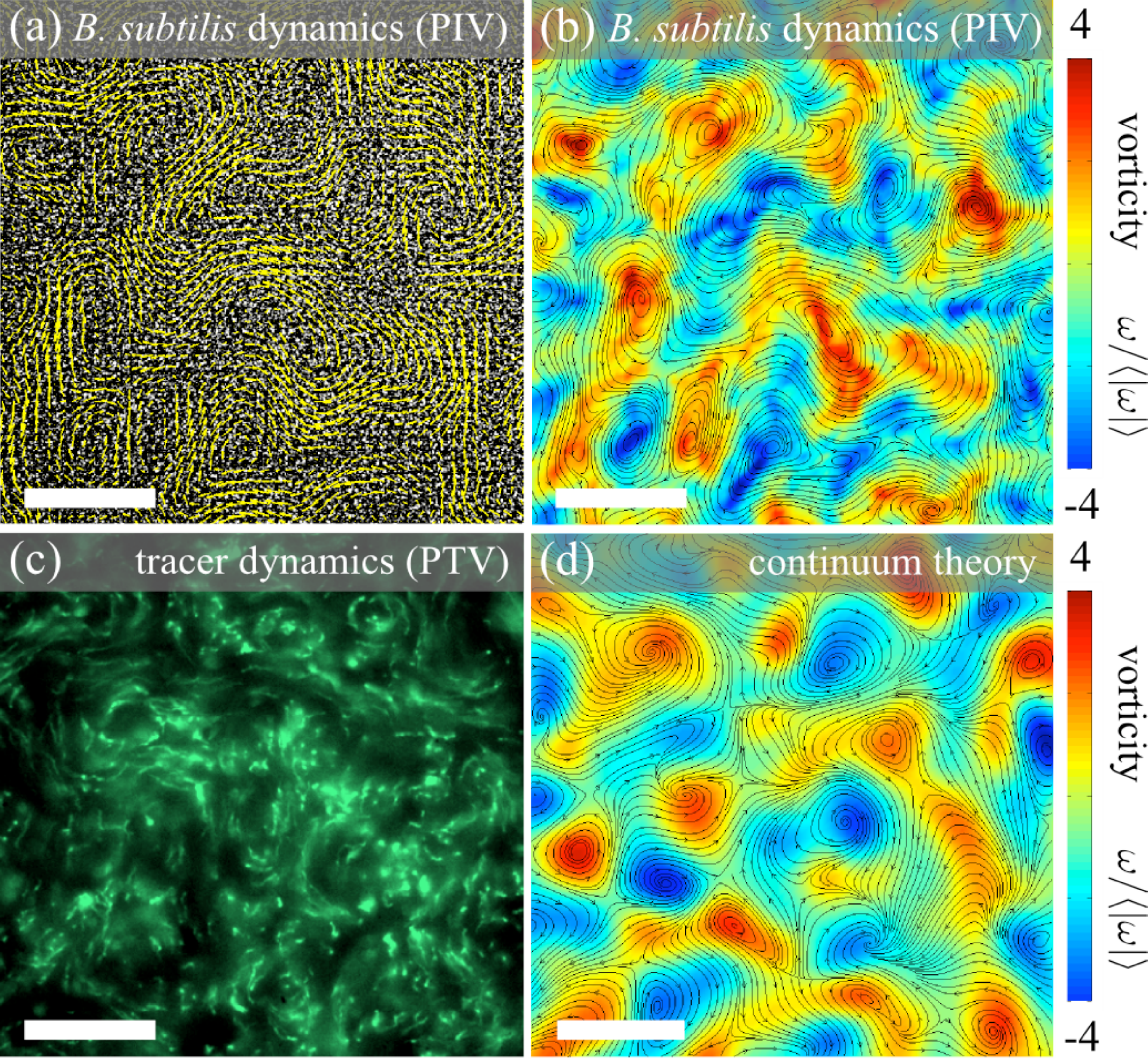}
\caption{(color online)  Flow fields from experiments and simulations~\cite{Supp}.
(a)~Very dense homogeneous suspension of {\it B. subtilis} overlaid with the PIV flow field showing 
collective bacterial dynamics.   
Longest arrows correspond to velocity of $30$~$\mu$m/s. (b)~Streamlines and normalized vorticity field 
determined from PIV data in~(a). 
(c)~Turbulent \lq Lagrangian\rq\space flow  of fluorescent tracer particles (false-color) in the same suspension, 
obtained by integrating emission signals over~1.5~s. 
(d)~Partial snapshot of a 2D slice from a 3D simulation of the continuum model 
(parameters in Table~\ref{t:sim_tab}). Scale bars 70~$\mu$m.
\label{f:ExpSim}
}
\end{figure}

\begin{figure}[t]
\center
\includegraphics[width=\columnwidth]{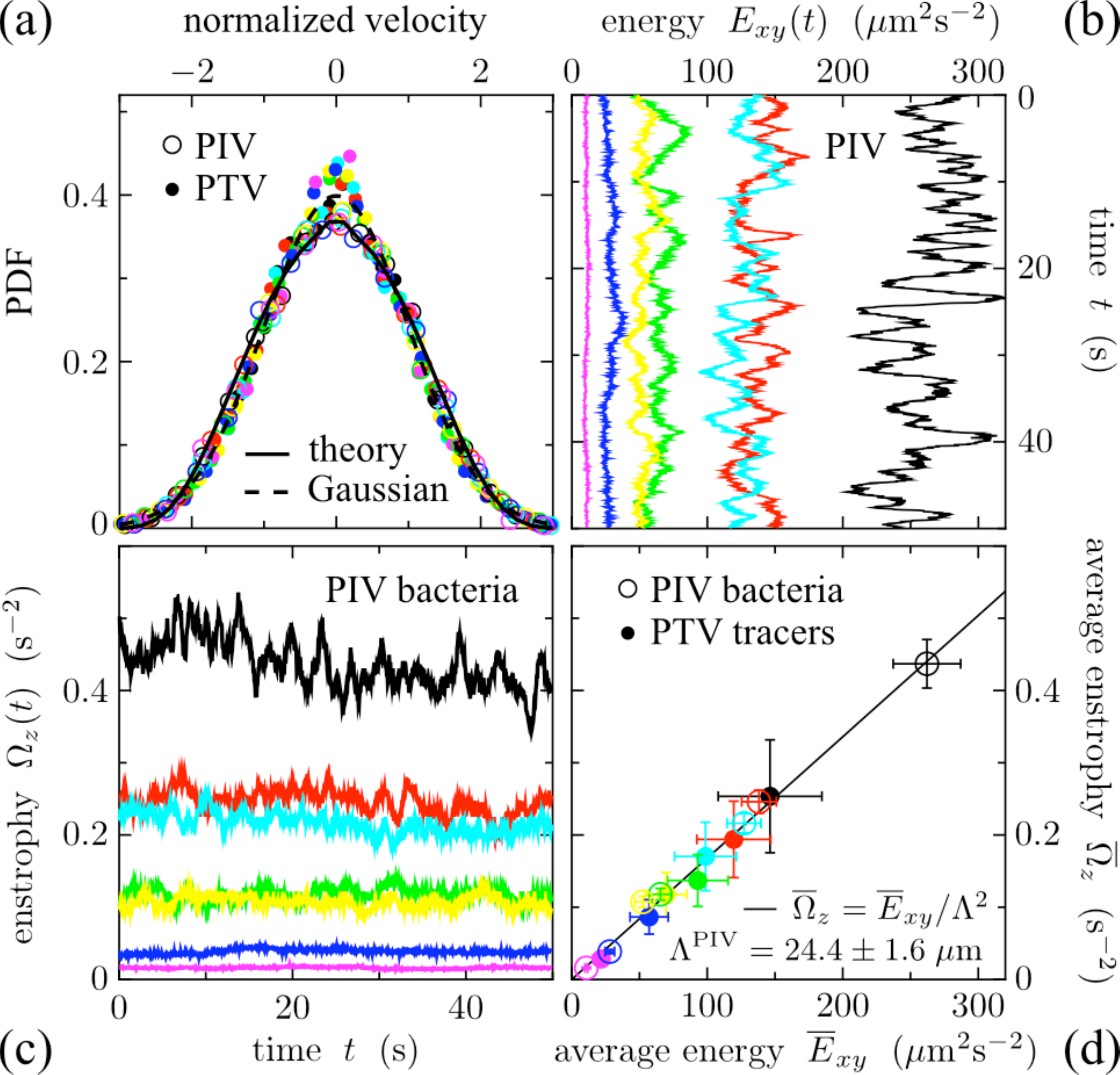}
\caption{\label{f:Ekin} (color online) 
Experimental results  for bacterial and medium flows, color-coded for activity level. 
(a)~PDFs of the Cartesian in-plane velocity components, normalized by their mean values and standard deviations, 
are approximately Gaussian (dashed) for both tracers and bacteria, with observable systematic deviations. 
The bacterial flow PDFs show slight broadening due to active swimming, which is well-reproduced by the 
model~\eqref{e:conti-model}. By contrast, the PTV distributions exhibit higher peaks at small velocities due to 
accumulation of tracers near vortex centers. (b,c)~Mean kinetic energy and enstrophy of the in-plane bacterial flow 
components  show moderate temporal fluctuations during the data aquisition period, very similar to corresponding 
PTV data (not shown). (d)~The time-averaged enstrophy scales linearly with the time-averaged energy. Open circles 
are averages of the curves in (b), (c). Errorbars indicate standard deviations.
}
\end{figure}

\begin{figure}[b]
\center
\includegraphics[width=\columnwidth]{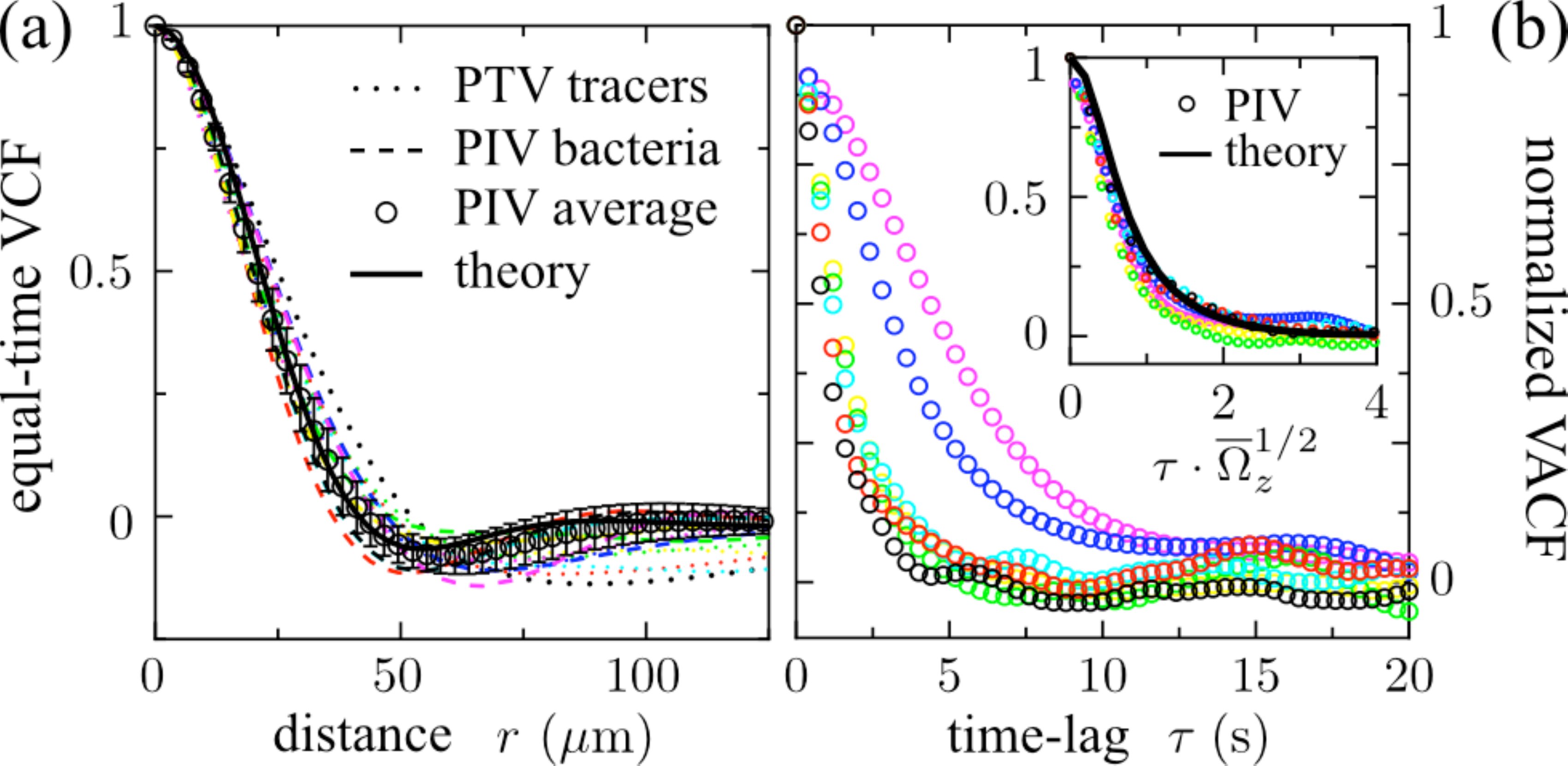}
\caption{\label{f:Corr} (color online) 
Correlation functions for solvent (PTV) and bacterial (PIV) flow at different energies and best-fit continuum theory (see Table~\ref{t:sim_tab}), using the same colors as in Fig. 2.  (a)~Both PIV and PTV data indicate a characteristic vortex radius $R_v\sim40\,\mu$m. The decay of spatial correlations at  small-$r$ depends only weakly on the activity level.
(b)~Velocity autocorrelation functions of the bacterial flow collapse when the time-lag $\tau$ is rescaled (inset) by the enstrophy time-scale  $\overline{\Go}_z^{-1/2}$. Since $\overline{\Go}_z\propto \overline{E}_{xy}$ (Fig.~\ref{f:Ekin}d), this implies that the higher the bacterial activity the shorter the flow memory.
}
\end{figure}

Wild-type strain 168 of {\it B. subtilis} has cigar-shaped cell bodies on average $0.8$  $\mu$m in diameter and $5$ 
$\mu$m long~\cite{2012Wensink}.  It was streaked on LB medium agar plates from frozen stocks. Colonies from these plates were 
used to inoculate overnight cultures in Terrific Broth (TB; Sigma), which were back-diluted 1:100 into 100$\,$ml of TB 
and grown to mid-logarithmic phase on an orbital shaker at 37$\,^{\circ}$C. These cultures were then concentrated
400$\times$ at $4,000\,g$ (final volume fraction $\sim 50$\%), and fluorescent microspheres (diameter 1$\,\mu$m, 
F-8816, Invitrogen) were added at a final concentration of $\sim$10$^9$~beads/ml. The resulting suspensions were 
loaded into polydimethylsiloxane (PDMS) microfluidic devices, consisting of a series of cylindrical chambers 
(radius 750$\,\mu$m, height 80$\,\mu$m), connected by thin channels \cite{2011DrescherEtAl,2012Wensink}.  
The inlet and outlet of the device were sealed with vacuum grease, and images were acquired in the $(xy)$-midplane of 
the chambers, $\approx$40$\,\mu$m above the bottom, using a Zeiss 40$\times$ (NA 1.3) oil immersion objective and a 
high-speed camera at 40$\,$fps (Fastcam SA-3, Photron). Movies were recorded in pairs for each field of view 
($768\times 800\,$pix;  1$\,$pix = $0.36\times0.36\,\mu$m$^2$), one with bright-field illumination and one with 
fluorescence excitation by a 633$\,$nm laser (B\&W Tek) at $\sim$20$\,$mW. These movies were taken immediately after 
each other with a $\sim$3$\,$min time lag between subsequent pairs. During the $\sim$10$\,$min imaging period for 
each device, the motility of {\it B. subtilis} cells decreased markedly due to oxygen depletion~\cite{2005Tuval_PNAS}. 
The  experimental setup yields 2D projected velocities of 3D suspension motion (Fig.~\ref{f:ExpSim}). Data were analyzed under 
the assumption that the flow structures are isotropic, as verified by test measurements at different distances from the 
chamber bottom.
Commercial particle tracking velocimetry (PIV) software (Dantec Flow Manager) was used to determine the bacterial flow velocity $(v_x,v_y)$ from 
bright-field images (Fig.~\ref{f:ExpSim}a,b), corrected for systematic pixel-locking errors~\cite{2011Cisneros_PRE}. 
Data shown in Figs.~\ref{f:Ekin} and~\ref{f:Corr} are based on $7$ movie segments (40$\,$fps, each 50$\,$s long)
corresponding  to 7 different  activity levels.

Global bacterial flows were quantified by the in-plane kinetic energy $E_{xy}(t)=\lan (v_x^2+v_y^2)/2 \ran$ 
and in-plane enstrophy $\Go_z(t)=\lan  \go_z^2/2\ran$, where $\go_z=\p_x v_y- \p_y v_x$ is the vertical 
component of vorticity and $\lan\,\cdot\,\ran$ is a spatial average. While $E_{xy}$ and $\Go_z$ fluctuate, 
their time averages $(\ovl{E}_{xy}, \ovl{\Go}_{z})$ are approximately constant during the $50$ s time interval 
used in the data analysis (Fig.~\ref{f:Ekin}b,c).  Over two orders of magnitude in energy  (Fig.~\ref{f:Ekin}d) 
we observe the linear scaling $\ovl{\Go}_z= \ovl{E}_{xy}/\Gl^2$, with $\Gl\approx 24\,\mu$m being roughly 
one half of the typical vortex radius.

Probability distribution functions (PDFs) of the in-plane bacterial velocity are approximately Gaussian, with
a slight broadening due to collective swimming~(Fig.~\ref{f:Ekin}a). The negative values of the equal-time spatial 
velocity correlation function (VCF; Fig.~\ref{f:Corr}a) indicate the existence of vortices 
\cite{2004DoEtAl} (Fig.~\ref{f:ExpSim}).  The VCF is remarkably robust with respect to changes
in the bacterial activity; in particular, the typical  vortex radius $R_v \sim40$\,$\mu$m, estimated from the first zero of the 
VCF,  depends only weakly on the kinetic energy.  This result is consistent with recent findings by Sokolov 
and Aranson~\cite{2012Sokolov} for free-standing films. The  vortex size in 3D is roughly five times larger than for 
quasi-2D turbulence in thin microfluidic chambers~\cite{2012Wensink}, where bacterial swimming and 
hydrodynamic interactions are suppressed by the nearby no-slip boundaries~\cite{2011DrescherEtAl,2012Spagnolie}. 
Unlike the spatial VCF,  the two-time velocity auto-correlation function (VACF) varies systematically with energy or 
vorticity (Fig.~\ref{f:Corr}b), but they collapse when plotted as 
functions of the dimensionless lag-parameter $\tau \Go_z^{1/2}$ (inset of Fig.~\ref{f:Corr}b), implying that the higher 
the activity the shorter the memory of the bacterial fluid. Generally, the statistics of 3D bacterial turbulence differ strongly 
from conventional 3D Navier-Stokes turbulence~\cite{2004Frisch,2006Bodenschatz}, as bacteria inject energy on the 
smallest scales, inducing an \lq upward\rq~energy cascade towards larger length scales.

We infer the flow of the solvent medium from particle tracking velocimetry (PTV) analysis of the fluorescence images, which only show the tracer particles,  
assuming that  they are passively advected.   Data shown in Figs.~\ref{f:Ekin} and~\ref{f:Corr}  are  based on 
7 movies (40~fps,  length 100~s) at different activities.  Trajectories of individual tracer particles were found with 
a custom algorithm which, depending on seeding density and tracer dynamics, was able to identify up to $10^4$
 in-plane tracks, the longest typically lasting $5-8\,$s. The effective sample size was insufficient to determine reliably the 
tracer VACFs, but did yield global flow properties, velocity histograms and equal-time VCFs. The velocity PDFs,  
calculated directly from individual tracer velocities, are approximately Gaussian with a peak at small velocities from 
tracer accumulation near the vortex centers  (Fig.~\ref{f:Ekin}a).

Estimates from PTV for the medium VCF and enstrophy were obtained by interpolating tracer velocities  
on  a~\mbox{$450\times450$~pix} subwindow in the center of the imaging plane  
using MATLAB's  Delaunay triangulation with a lattice spacing $\Gd=90\,\sqrt{\text{pix}/{N}_f}$, where ${N}_f$ is the 
mean number of tracers detected per frame.  The accuracy of this reconstruction procedure is controlled by the 
tracer concentration, which was kept low  to limit effects on the bacteria motion and to avoid tracking ambiguities 
(typically $N_f\in[ 47,144]$ for data shown in Figs.~\ref{f:Ekin} and~\ref{f:Corr}). As a result, the uncertainties 
for the PTV data are considerably larger than for PIV data (see Fig.~\ref{f:Ekin}d). The interpolated tracer flow fields 
were used to estimate the kinetic energy $E_{xy}$, enstrophy~$\Go_z$,  and spatial correlation functions of the 
in-plane medium flow components. In agreement with the PIV results for the bacterial flow, we find again a linear 
enstrophy-energy relation (Fig.~\ref{f:Ekin}d) and comparable vortex radii, using the first zero of VCF as an 
estimate (Fig.~\ref{f:Corr}a). We may therefore conclude that, at our very high bacterial concentrations, solvent and 
bacterial flow statistics become tightly linked.

We now examine how these data compare to predictions of a theory of active fluids introduced recently
\cite{2012Wensink,2012DuEtAl}. This minimal continuum model assumes that, at high concentrations,  the bacterial flow 
due to swimming and advection can be described by a single velocity field $\bs v(t,\bs x)$ and a pressure $p(t,\bs x)$. 
They obey the incompressibility condition $\nabla \cdot \boldsymbol v=0$  and   
 \begin{eqnarray}
 (\partial_t + \lambda_0 \boldsymbol v\cdot \nabla) \boldsymbol v
 &=&\notag
-\nabla p  +\lambda_1 \nabla \boldsymbol v^2 - \gb (\boldsymbol v^2-v_0^2)\boldsymbol v + 
\\&&
\Gamma_0 \nabla^2 \boldsymbol v -\Gamma_2(\nabla^2)^2    \boldsymbol v.
\label{e:conti-model}
 \end{eqnarray}
Equation~\eqref{e:conti-model} extends the incompressible Toner-Tu
theory~\cite{2005ToTuRa,2010Ramaswamy,1998TonerTu_PRE} with a fourth order term as in the 
Swift-Hohenberg equation~\cite{1977SwiftHohenberg}.  The parameter $\gl_0$ describes advection and nematic 
interactions, and $\gl_1$ an active pressure contribution~\cite{2012DuEtAl}. For pusher 
swimmers~\cite{2011DrescherEtAl}  like \emph{B. subtilis}, general considerations of  hydrodynamic~\cite{2010Pedley}  
and nematic stresses~\cite{2008BaMa,2012DuEtAl} suggest that \mbox{$\lambda_0\ge 1$} and 
{$ \lambda_1\simeq (\gl_0-1)/3\ge 0$} in 3D. The $(\beta, v_0)$-terms correspond to a quartic Landau-type velocity
potential~\cite{2005ToTuRa,2010Ramaswamy,1998TonerTu_PRE} and are physically motivated by the observation of 
extended jet-like streaming regions in \emph{B.~subtilis} suspensions at intermediate
concentrations~\cite{2011Cisneros_PRE}.  The parameter $v_0$ defines the collective speed that would be achieved 
if all bacteria were to move in the same direction. When $\beta\ne0$ the model does not conserve momentum or energy, 
as it describes exclusively the bacterial flow component, which may exchange energy and momentum with the solvent. 
The nonlocal $(\Gc_0,\Gc_2)$-terms encode passive and active stresses due to hydrodynamic and steric interactions.  
For $\gl_0=1$, $\gl_1= \beta=\Gc_2=0 $ and $\Gc_0>0$, the model reduces to the incompressible Navier-Stokes equation. 
A detailed stability analysis~\cite{2012DuEtAl} shows that when $\gl_0\ne 0$, $\beta>0$, $v_0>0$,  $\Gc_2>0$ 
and $\Gc_0<0$ this is one of the simplest vector models to describe phenomenologically the formation of jets and 
turbulent vortices  in quasi-incompressible active suspensions.   Very recently, the 2D version of Eq.~\eqref{e:conti-model} 
has been shown to provide a quantitative mean field description of bacterial meso-scale turbulence in 
quasi-2D suspensions~\cite{2012Wensink}.  Its applicability to the physically more relevant 3D case is first
explored here.

\begin{figure}[t]
\center
\includegraphics[width=\columnwidth ]{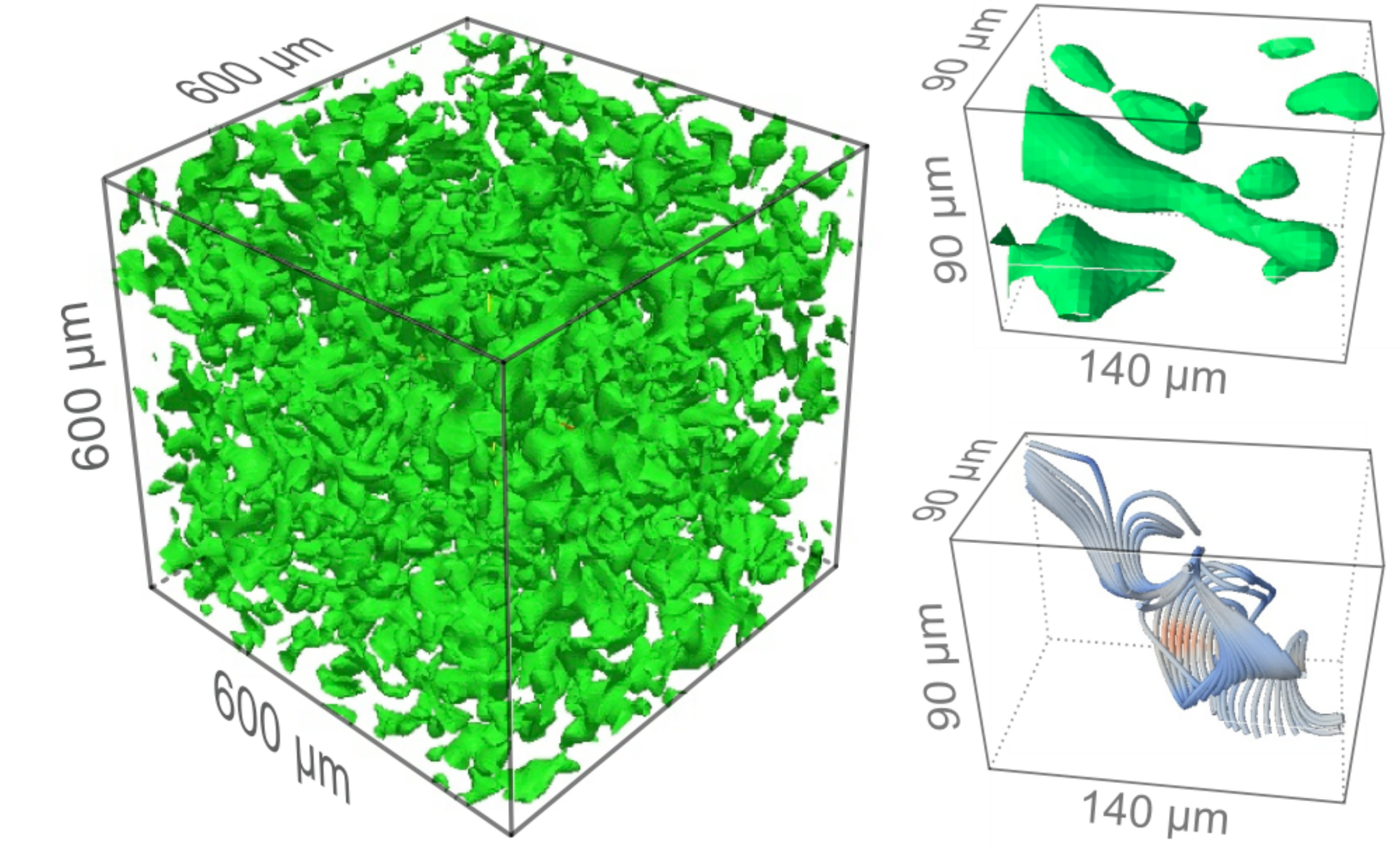}
\caption{\label{f:3d}  (color online) 
Isoenergy surfaces ($E=1.7v_0^2$) and selected stream tubes from the best-fit 3D simulation (visualized with ParaView)  indicate a typical vortex length scale and extended band-like regions corresponding to co-aligned bacterial jets.   See Supplemental Material \cite{Supp} for a movie.
}
\end{figure}

We simulated Eq.~\eqref{e:conti-model} in 3D with periodic boundary conditions using a pseudospectral  
operator-splitting algorithm~\cite{Orszag,Pedrosa} and a pressure correction subroutine to 
ensure incompressibility~\cite{2012Wensink,2012DuEtAl}.  Simulation grids ranged from  $128^3$ lattice points for 
parameter pre-screening to $256^3$ for statistical analysis. Numerical stability of the solver was verified 
for a wide range of parameters and space-time discretizations.   All simulations were initiated with randomly chosen velocities. 
Figure \ref{f:3d} shows structure-formation in a typical simulation domain.

Since in 3D we have $\lambda_1\simeq(\gl_0-1)/3$~\cite{2012DuEtAl},  Eq.~\eqref{e:conti-model} has essentially five free parameters $(\gl_0,\beta,v_0,\Gamma_0,\Gamma_2)$. Two of those can be eliminated  by choice of appropriate length and time units. We adopt a natural unit system such that  the vortex wave-length scale $\Gl_\Gc=2\pi \sqrt{\Gc_2/(-\Gc_0)}=2\pi$ and $v_0=1$. In our simulations,  the box length is fixed as $L=12\Gl_\Gc$, corresponding to approximately twice the experimental field of view,  and the time step as $\Delta t = 0.05 \Gl_\Gc/(2\pi v_0)$.  To estimate the three remaining parameters $(\gl_0,\beta,\Gamma_0)$, we note that $\Gc_0$ and $\Gamma_2$ define a 
typical vortex speed $V_\Gc=\sqrt{-\Gc_0^3/\Gc_2}$.  In the turbulent regime, it is plausible that $V_\Gc$ is smaller than but close to $v_0$, i.e. $V_\Gc=\zeta v_0$ where $\zeta\lesssim 1$. Furthermore, for pushers, the dimensionless parameter $\lambda_0$ 
should be larger than~$1$, but smaller than for quasi-2D suspensions~\cite{2012Wensink}, since nematic (steric) stresses 
can be more easily avoided in 3D;  we infer $\gl_0\sim 2$. Finally, the acceleration time scale 
$\tau_0=(\beta v_0^2)^{-1}$ should be of the order of the vortex time-scale $\Lambda_\Gamma/V_\Gamma$.  Using 
these estimates as initial values in a systematic parameter scan, and by comparing with the bacterial PIV data, we 
obtained the best-fit parameters in Table~\ref{t:sim_tab}. Generally, the VCFs and VACFs  respond sensitively to 
parameter variations in the simulations, suggesting that the estimates in Table~\ref{t:sim_tab} are accurate within 10-15\% for quasi-incompressible \textit{B. subtilis} suspensions. As an independent cross-check, we computed $\Lambda=(\overline{E}_{xy}/\Omega_z)^{1/2}$ from the best-fit simulation using $\Lambda_\Gamma\sim 50\,\mu$m and found  $\Lambda\sim 29\,\mu$m  which compares well  with the experimental PIV value in Fig.~\ref{f:Ekin}d. We stress that the conserved form of the bacterial velocity PDFs 
(Fig.~\ref{f:Ekin}a), VCFs, and VACFs (Fig.~\ref{f:Corr}) implies that all our experiments can be fitted by a single set 
of rescaled parameters $(\gl_0,\beta,\Gamma_0)$, as it suffices to adjust the physical values of $v_0$ and $\Gl_\Gc$ to match the kinetic energy and vortex length at a given bacterial activity level.  As evident from the flow patterns in Fig.~\ref{f:ExpSim} and from the 
solid curves in Figs.~\ref{f:Ekin}a  and~\ref{f:Corr},  the best-fit parameters yield good qualitative and quantitative 
agreement with the experiments. 

\begin{table}[t]
\centering
\begin{tabular}{c|c|c}
model parameter  &in rescaled units  & in physical units  \\
\hline
$\Gl_\Gc = 2\pi \sqrt{\Gc_2/(-\Gc_0)}$ &  	$2\pi$ 	&	$\sim 50\;\mu$m \\
$v_0$	&  	1               & 	$3 - 22\;\mu$m/s        \\
\hline
$ \gl_0$ & $  1.7 $ & $  1.7 $  \\
$V_\Gc=\sqrt{-\Gc_0^3/\Gc_2}$  & 0.9		&  	$0.9 v_0$\\
$\gb $ &  0.1 &  $ 1.3\times 10^{-2}(v_0\,\mu$m$)^{-1}$  \\
\hline
$\overline{E}_{xy}$ & 0.54 & $0.54v_0^2$ 
\end{tabular}
\caption{Parameters of the best-fit continuum model. To match a specific experiment, one must merely adjust the physical value of $v_0$ by equating $\overline{E}_{xy}=0.54v_0^2$ to the corresponding kinetic energy value in Fig.~\ref{f:Ekin}d.
\label{t:sim_tab}}
\end{table}

For incompressible \lq passive\rq~fluids, that are governed by the Navier-Stokes equations transport parameters 
have of course been measured for a wide range of materials~\cite{ViscosityHandbook}. In contrast, quantitative 
theories of even the simplest active fluids have been lacking. We have shown here that the minimal 
fourth-order vector model~\cite{2012DuEtAl,2012Wensink} in 
Eq.~\eqref{e:conti-model} reproduces the main statistical features of self-sustained 3D bulk turbulence in 
concentrated bacterial suspensions, suggesting that this theory is a viable candidate for the quantitative 
description of incompressible active fluids. 
Due to the close correlation between bacterial and medium (tracer) flow observed in our experiments, we expect 
that this generic model will be useful in a wide range of future applications, in particular for predicting the effects of 
confining geometries on collective microbial dynamics ~\cite{2012Woodhouse,2013Ravnik_PRL} and for understanding 
the anomalous viscosities of active fluids~\cite{2009SoAr,2009Rafai}. 

The authors would like to thank Sabine Klapp, Hartmut L\"owen,  Cristina Marchetti, Lutz Schimanksy-Geier, Holger Stark, Hugo Wioland, Francis Woodhouse and Julia Yeomans for helpful discussions. 
This work was supported by the Deutsche Forschungsgemeinschaft, GRK1558 (M.B. and S.H.), the Human Frontier 
Sciences Program (K.D.), and the European Research Council, Advanced Investigator Grant 247333 (J.D. and R.E.G.). 
J.D. and S.H. contributed equally to the paper and are joint first authors.

\bibliography{refs}


\end{document}